\documentclass{aa}
\usepackage{graphicx}
\usepackage{natbib}
\usepackage{color}
\usepackage{fancyvrb} 
\usepackage{amssymb}
\usepackage{bm,url,times}
\graphicspath{{./fig/}{./ps/}}

\newcommand{\Eq}[1]{Eq.~(\ref{#1})}
\newcommand{\Fig}[1]{Fig.~\ref{#1}}
\newcommand{\Tab}[1]{Table~\ref{#1}}
\newcommand{\dd}{{\rm d} {}}
\newcommand{\EQ}{\begin{equation}}
\newcommand{\EN}{\end{equation}}
\newcommand{\EQA}{\begin{eqnarray}}
\newcommand{\ENA}{\end{eqnarray}}
\newcommand{\Eqs}[2]{Eqs.~(\ref{#1}) and~(\ref{#2})}

\title{The radial gradient of the near-surface shear layer of the Sun}
                                   
\author{A. Barekat\inst{1} \and J. Schou\inst{1} \and L. Gizon\inst{1,2}}

\date{\today}

\institute{
Max-Planck-Institut f\"ur Sonnensystemforschung, Justus-von-Liebig-Weg
3, 37077 G\"ottingen, Germany
\and
Institut f\"ur Astrophysik, Georg-August-Universit\"at G\"ottingen,
37077 G\"ottingen, Germany
}
\begin{document}

\abstract
{
Helioseismology has provided unprecedented information about the internal
rotation of the Sun. One of the important achievements was the discovery of
two radial shear layers: one near the bottom of the convection
zone (the tachocline) and one near the surface. These
shear layers may be important ingredients for explaining the magnetic
cycle of the Sun.  
}
{
We measure the logarithmic radial gradient of the rotation rate
($\dd\ln\Omega/\dd\ln r$) near the surface of the Sun using 15 years
of f~mode rotational frequency splittings from the Michelson Doppler
Imager (MDI) and four years of data from the Helioseismic and Magnetic Imager (HMI).
}
{
We model the angular velocity of the Sun in the upper $\sim 10$~Mm
as changing linearly with depth and use a multiplicative optimally
localized averaging inversion to infer the gradient of the
rotation rate as a function of latitude.
}
{
Both the MDI and HMI data show that $\dd\ln\Omega/\dd\ln r$ is close
to $-1$ from the equator to  
60$^{\circ}$ latitude and stays negative up to 75$^{\circ}$
latitude. However, the value of the gradient is different for MDI and HMI  
for latitudes above $60^{\circ}$. Additionally, there is a significant
difference between the value of $\dd\ln\Omega/\dd\ln r$ using an older
and recently reprocessed MDI data for latitudes above $30^\circ$.  
}
{We could reliably infer the value of $\dd\ln\Omega/\dd\ln r$ up to
  60$^{\circ}$, but not above this latitude, which will hopefully constrain
  theories of the near-surface shear layer and dynamo. 
Furthermore, the recently reprocessed MDI splitting data are more
  reliable than the older versions which contained clear systematic
  errors in the high degree f modes.   
}


\maketitle

\section{Introduction}
\label{introduction}

Helioseismology has had a significant impact on our understanding of
the internal structure and dynamics of the Sun. One of the most important
results has been the inference of the rotation profile \citep{JS98}. 
Two shear layers have been identified, one located near the base of
the convection zone \citep{JCD88,Brown89}, known as the tachocline
\citep{Spiegel92}, 
and one in the upper 35 Mm, the near-surface shear layer
\citep[NSSL,][]{Thompson96}. Explaining the current picture of the
internal rotation profile in theoretical terms is a major challenge.
\citep{KiRu93,KiRu05}. 

The rotation profile in general and shear layers in particular may
play a crucial role for the solar dynamo
\citep[e.g., ][]{AxelK05, Charb10}.  
This led to further investigation of the NSSL using helioseismic
measurements (\citealt{Basu99}; \citealt{CT02}, hereafter CT; \citealt{Howe06}; \citealt{Zaatri09}) and its role in dynamo
theory \citep{Dikpati02, Mason02,Axel05,Kapyla06}. 
The logarithmic radial gradient of the rotation  
rate ($ \dd \ln \Omega / \dd \ln r$) evaluated at the surface was
measured by CT using f modes. They used 23
data sets (each from 72-day time series) of 18 odd
$a$-coefficients from the Medium-$l$ program \citep{Scher95} of the
Michelson Doppler Imager (MDI) onboard the Solar and Heliospheric
Observatory (SOHO) covering the years 1996 through 2001.  
Their main result was that $\dd \ln \Omega / \dd \ln r \sim -1$ up to
$30^\circ$ latitude,
reverses sign around $55^{\circ}$ latitude and stays positive at
higher latitudes.  
However, they also noted that there are indications of systematic errors mostly
affecting high latitudes.
We address this issue by analyzing splittings from 
MDI and the Helioseismic and Magnetic Imager
\citep[HMI;][]{Schou12} onboard the Solar Dynamics Observatory (SDO).
\section{Observations}
\label{OBS}
Thousands of oscillation mode frequencies $\nu_{nlm}$ can be measured
on the Sun, where $n$, $l$, and $m$ are the radial order, the spherical
harmonic degree, and the azimuthal order, respectively. 
The mode frequencies  $\nu_{nlm}$  are expanded using so-called
$a$-coefficients  \citep{JCT94}
\EQ
\nu_{nlm}= \nu_{nl} + \sum_{j=1}^{j=36}a_{nl,j}\mathcal{P}_j^{(l)}(m),
\EN
where $\nu_{nl}$ is the mean multiplet frequency and $\mathcal{P}_j^{(l)}$ are
orthogonal polynomials of degree $j$. This work considers
only f modes, for which $n=0$, and so we suppress $n$ in the following.
We use two sets of $a$-coefficients. The first is from
the MDI Medium-$l$ 
program and contains 74 sets of splittings from
independent 72-day time 
series (Larson \& Schou in prep.). These data cover about 15 years from
1996 May 1 to 2011 April 24,  
except for 1998 from July 2 to October 17 and 1998 December 23 to 1999
February 2 due to  
technical problems with SOHO.
The second set is from HMI and contains 20 sets of
splittings from consecutive 72-day time series (Larson \& Schou in prep.),
covering four years of  
observation from 2010 April 30 to 2014 April 8. 
Additionally, in order to compare our results with the results
obtained by CT, we also use older version of the MDI data.
The differences between these versions come from various
improvements to the analysis, as described in \cite{TL09} and (Larson
\& Schou in prep.). 
We refer to the older version as ``old MDI'' and to the latest ``new MDI''.
The f modes we use cover the range $117\leq l \leq 300$ for MDI and 
$123\leq l \leq 300$ for HMI. We note that the number of available
modes changes with time because of noise. 

\section{Analysis of f mode data}
\label{analysis}
The odd $a$-coefficients are related to the angular velocity $\Omega$
by
\EQ
2\pi a_{l,2s+1} = \int_0^1  \int_{-1}^1 \; K_{ls}(r,u)\Omega(r,u)\dd u\dd r ,
\label{a-coff}
\EN
where the kernels $K_{ls}$ are known functions, $u=\cos\theta$,
$\theta$ is the co-latitude, and $r$ 
is the distance to the center of the Sun divided by the photospheric radius.
Using the results of \cite{Pij97}, one can show that the kernels can
be separated 
in the variables $r$ and $u$,
\EQ
K_{ls}(r,u)=F_{ls}(r)G_s(u),
\label{klsf}
\EN
where the functions $F_{ls}$ and $G_s$ are the radial and latitudinal
parts of the kernels. The function $F_{ls}$ is
\EQ
F_{ls}(r)=\Big[F_{l,1}(r)- F_{l,2}(r) (2s+2)(2s+1)/ 2 \Big] v_{l,2s+1},
\EN
where $F_{l,1}$, $ F_{l,2}$ and $ v_{l,2s+1}$ are given by
\EQA
F_{l,1}(r)&=& \rho(r) r^2\left[\xi_l^2(r)- 2 \xi_l(r)\eta_l(r) /L +\eta_l^2(r)\right]/I_{l},\\
F_{l,2}(r) &=& \rho(r) r^2 \eta_l^2(r)/(L^2 I_{l}),\\
v_{l,2s+1}&=&{ (-1)^s \over l} {(2l+1)!(2s+2)!(l+s+1)!\over
s!(s+1)!(l-s-1)!(2l+2s+2)!} .
\ENA
In the above equations $\rho$ is the density, $L=\sqrt{l(l+1)}$, $\xi$ and $\eta$ 
are the radial and horizontal displacement eigenfunctions as defined
by \cite{Pij97}, and  $I_l=\int_0^1\rho(r) r^2 \left[\xi_l^2(r)+ \eta_l^2(r)\right]\dd r$.
The latitudinal part of the kernels is given by
\EQ
G_s(u)=- {(4s+3) \over 2(2s+2)(2s+1)}(1-u)^{1/2} P_{2s+1}^1(u),
\EN
where $P_{2s+1}^1$ are associated Legendre polynomials of
degree $2s+1$ and order one.
As seen later, the form of \Eq{klsf} is useful in that the
latitudinal part of the kernels is independent of $l$.

\begin{figure}[t!]
\begin{center}
\includegraphics[width=\columnwidth]{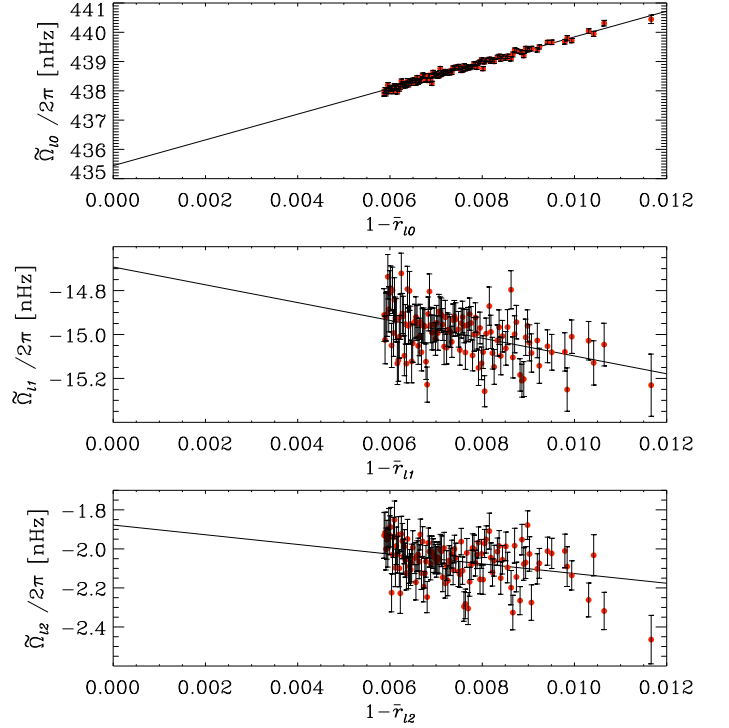}
\end{center}\caption[]{$\widetilde{\Omega}_{ls}/2\pi$ versus
  $(1-\overline{r}_{ls})$ for $s=0,1,$ and 2 from top to bottom
  for the HMI data set starting on 2014
  January 27. The error bars are 1$\sigma$. 
}\label{fit}\end{figure}

We use f modes to calculate
$\dd\ln\Omega/\dd\ln r$
close to the surface of the Sun in several
steps. In the first step, we assume that the rotation rate changes
linearly with depth at each latitude
\EQ
\Omega(r,u)=\Omega_0(u)+(1-r)\Omega_1(u),
\label{omega}
\EN
where $\Omega_1$ is the slope and $\Omega_0$ is the value of the
rotation rate at the surface.
Combining \Eq{omega} with \Eqs{a-coff}{klsf} we obtain
\EQ
\widetilde{\Omega}_{ls} \equiv {2\pi a_{l,2s+1}\over \beta_{ls} } =
\langle\Omega_0\rangle_s + (1-\overline{r}_{ls}) \langle
\Omega_1 \rangle_s,
\label{fittt}
\EN
where $\beta_{ls}=\int_0^{1} F_{ls}(r)\dd r$ 
and $\overline{r}_{ls}=\beta^{-1}_{ls}\int_0^{1} F_{ls}(r)r\dd r 
$ is the center of gravity of  $F_{ls}$.
The quantities $\langle\Omega_0\rangle_s$ and
$\langle\Omega_1\rangle_s$ are the latitudinal averages  
\EQA
\langle\Omega_0\rangle_s &=&\int_{-1}^1G_s(u)\Omega_0(u)\dd
u, \label{omega0}\\
\langle \Omega_1 \rangle_s&=&\int_{-1}^1G_s(u)\Omega_1(u)\dd u.
\label{omega1}
\ENA
By performing an error weighted linear least squares fit of $\widetilde{\Omega}_{ls}$ versus
$(1-\overline{r}_{ls})$ 
we can estimate  $\langle\Omega_0\rangle_s$ and $\langle\Omega_1\rangle_s$.
This procedure is applied for all $s$ with $0 \leq s \leq 17$ for each
individual  72-day data set. 
To illustrate this, \Fig{fit} shows $\widetilde{\Omega}_{ls}/2\pi$ as
a function of $(1-\overline{r}_{ls})$ for $s=0, 1,$ and $2$ for
one time period.
We note that the values of $(1-\overline{r}_{ls})$ correspond to a
depth range of about $4.5\,-\,8.4$ ~Mm, and that the kernels have a
significant extent in depth. 
Our estimates of $\Omega_0$ at the surface are thus in effect
extrapolations and the values of $\Omega_1$ are averages,
both estimated from roughly the outer 10~Mm. 

Next we invert $\langle\Omega_0\rangle_s$ and
$\langle\Omega_1\rangle_s$ to obtain estimates
$\overline{\Omega}_0(u_0)$ and $\overline{\Omega}_1(u_0)$ of
${\Omega}_0(u_0)$ and ${\Omega}_1(u_0)$, where $u_0$ is the target
point for the inversion. 
Following \cite{JS99}, we use a
multiplicative optimally localized averaging inversion method with a
trade-off parameter $\mu=0$. 
This implies that the averaging kernels for 
$\langle\Omega_0\rangle_s$ and $\langle\Omega_1\rangle_s$
are the same as those shown in Figure 4 of \cite{JS99}.

Finally, we obtain an estimate of the surface value of the logarithmic
radial gradient of the angular velocity at each $u_0$ as
\EQ
\left ( { \dd\ln\Omega \over \dd\ln r }\right ) (r=1,u=u_0)  \approx -
{\overline{\Omega}_1(u_0)\over \overline{\Omega}_0(u_0)}.
\label{dodr}
\EN

\section{Results}
\label{result}

\begin{figure}[t!]
\begin{center}
\includegraphics[width=\columnwidth]{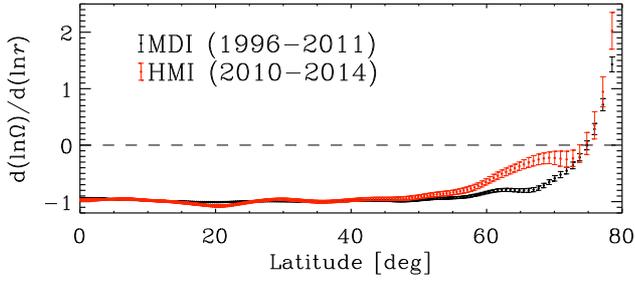}
\end{center}\caption[]{Time average of
$\dd\ln\Omega/\dd\ln r$ versus target latitude, 
obtained from 15 years (1996-2011) of MDI data (black dots) and 4
years (2010-2014) of HMI data (red dots). The error bars are $1\sigma$.
}\label{comp}\end{figure} 

\begin{table}[!t]\caption{Selected values of $\dd\ln\Omega/\dd\ln r$ from \Fig{comp}.
}\vspace{12pt}\centerline{\begin{tabular}{ccc}                            
\hline \hline
Latitude [deg] & MDI &HMI \\
\hline
0  &  $-0.939\pm 0.009$ & $-0.97\pm 0.02$\\ 
10 &  $-0.981\pm 0.007$ & $-0.98\pm 0.02$\\ 
20 &  $-1.009\pm 0.007$ & $-1.08\pm 0.02$\\ 
30 &  $-0.992\pm 0.009$ & $-0.96\pm 0.02$\\  
40 &  $-0.986\pm 0.011$ & $-0.97\pm 0.03$\\
50 &  $-0.974\pm 0.014$ & $-0.92\pm 0.03$\\
60 &  $-0.841\pm 0.022$ & $-0.65\pm 0.05$\\
70 &  $-0.588\pm 0.048$ & $-0.23\pm0.12$\\
\hline              
\label{LRG-target}\end{tabular}}
\end{table}

\begin{figure}[t!]
\begin{center}
\includegraphics[width=\columnwidth]{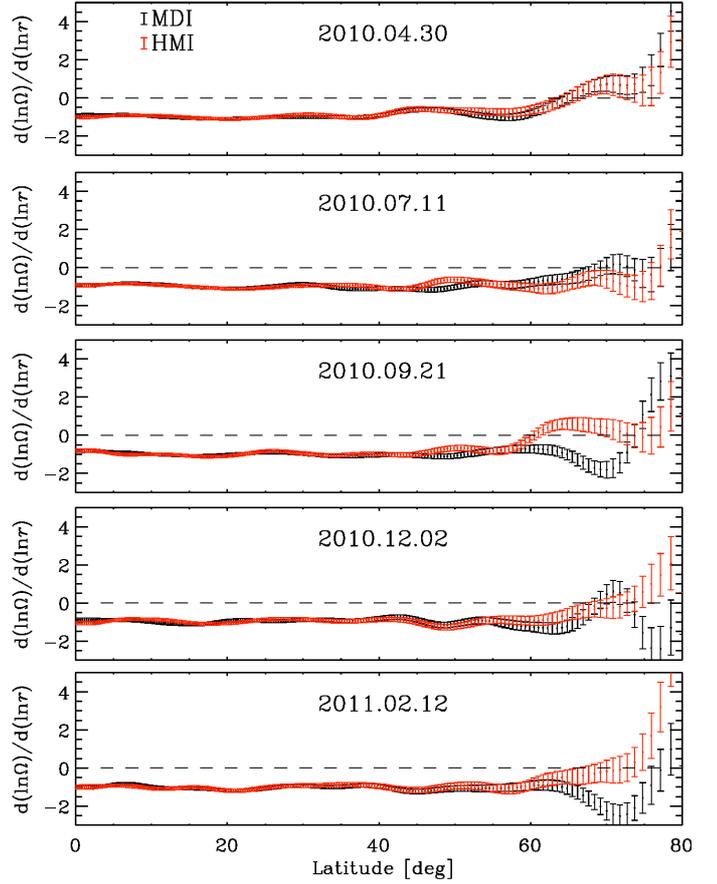}
\end{center}\caption[]{
Comparison of 
$\dd\ln\Omega/\dd\ln r$ versus target latitude for MDI (black dots) and HMI (red dots)
from the five common 72-day time series
(indicated by the nominal beginning dates).
Error bars are 1-$\sigma$. 
}\label{comparison}\end{figure}

In \Fig{comp} we plot the estimates of
$\dd\ln\Omega/\dd\ln r$
as a function of target latitude ($\arcsin u_0$) averaged over time for 15 years of new MDI data and
4 years of HMI data; in \Tab{LRG-target} we give the results.
The results are similar and very close to $-1$ from the equator
to $\sim 60^{\circ}$ latitude, while above $60^\circ$ they diverge.
The differences at high latitudes could be due to either systematic
errors or a solar cycle effect (the data sets cover different
parts of the solar cycle). To investigate this discrepancy,
\Fig{comparison} shows the results of applying our method to
 the HMI
and MDI data sets from the five common 72-day periods between 2010
April 30 and 2011 April 24.
The results are consistent up to $\sim 60^\circ$ within
2-$\sigma$, but show 
significant inconsistencies at higher latitudes.
An analysis using only the common modes and the
average errors does not significantly reduce this high latitude discrepancy.
This indicates that there are systematic errors in at least one of
the data sets, as opposed to only  differences in the mode
 coverage or error estimates.
The source of the systematic errors is unknown, but could be related to
inaccurate estimates of the optical distortion of the instruments or
similar geometric 
errors (Larson \& Schou in prep.). Another possible source is the
different duty cycles. For example, the last three
data sets for MDI had duty cycles of 
88\%, 73\%, and 81\%, while the corresponding HMI duty cycles were
97\%, 99\%, and 96\%. In either case we conclude that the results
above $\sim 60^\circ$ should be treated with caution. 

The results presented here are significantly different
from those obtained by CT. They found that
$\dd\ln\Omega/\dd\ln r$ is close to $-1$ from the equator to $30^\circ$
latitude, while our result shows this up to $60^\circ$ latitude. 
They also found that their results changed significantly if they
restricted the degree range. To investigate the origin of these
differences we examine the effects of each of the differences
between their data and analysis and ours. 

\begin{figure}[t!]
\begin{center}
\includegraphics[width=0.9\columnwidth]{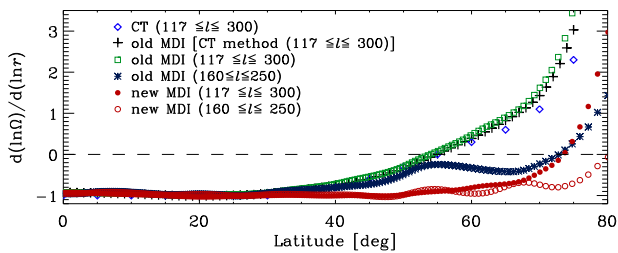}
\end{center}\caption[]{
Estimates of $\dd\ln\Omega/\dd\ln r$ versus target latitude obtained from 23 MDI data sets
using various methods.
Blue diamonds show the values measured from Fig. 4 of CT,
while black pluses
show the results of changing the data sets and averaging, as described
in the text. 
Green squares and dark blue stars show the results of our analysis of
the old MDI data 
for the full and restricted modes, respectively.
Filled and open red circles show the corresponding results for the new MDI data.
}\label{CT-comp}\end{figure}

First, we compare the results of applying our method and theirs to the 23
time periods they used (covering the period 1996 May 1 to
2001 April 4).
\cite{CT02} first made an error weighted time average of an older version of the MDI data and then
applied their Eq.~(9). If we repeat this procedure on the same data sets we obtain
results visually identical to theirs.
The difference between the data sets used by CT and old MDI is that a few modes were accidentally removed
from the older set.
We then changed the processing order to first apply their Eq.~(9)
to old MDI and then make an unweighted time average.
As shown in \Fig{CT-comp}, this results in minor differences at high latitude and an analysis
applying each change separately shows that only the change from weighted fits to unweighted fits
leads to a noticable difference.

We then restricted the old MDI mode set to $160\leqslant l \leqslant 250$.
 As shown in \Fig{CT-comp}, this results in large changes above $\sim
 50^\circ$, in agreement with what CT found.
This indicates that the linear model of the rotation rate (as given by
\Eq{omega}) is incorrect or that there are systematic errors.

Finally, we apply our method to the old MDI and new MDI data sets.
As can be seen in \Fig{CT-comp} we see a significant
difference above $30^\circ$ latitude.
The result using the new MDI
data does not show any change of the sign up to $\sim 55^\circ$ latitude
and is $\sim -1$ up to $60^\circ$ latitude. 
The results using the new MDI data sets also show good agreement between the
results of the complete and restricted mode sets up to almost
$70^\circ$ latitude, indicating that the model of
linear change of the angular velocity with depth represents those
data better than the old MDI data. 

\begin{figure}[t!]
\begin{center}
\includegraphics[width=0.95\columnwidth]{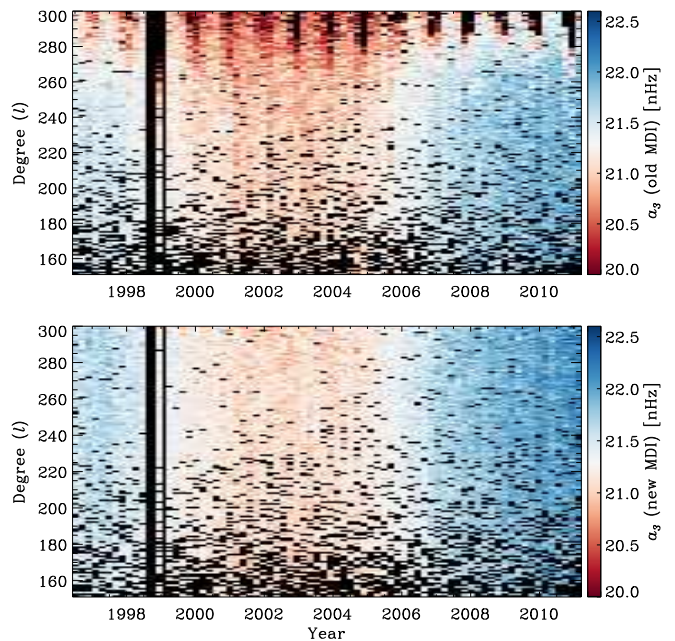}
\end{center}
\caption[]{$a_3$ for old MDI (upper panel) and new MDI
  (lower panel) for $150\leqslant l \leqslant 300 $ over time. Black
  shows missing modes.
  For clarity a few old MDI values below $20$~nHz were set to $20$~nHz.
}\label{mdicomp}\end{figure}

As almost all the differences between the results obtained by CT and
ours come from the differences between old and new MDI, we
compare the $a$-coefficients directly.
As an example, \Fig{mdicomp} shows $a_3$ for the
modes with $150\leqslant l \leqslant 300 $ for all 74 periods.
The main differences
between new and old MDI appear for $l>270$.
In the new MDI data most of the missing modes (shown in black) in the old MDI data
are recovered and the yearly oscillatory pattern disappears.
These differences clearly show that the old MDI data have 
significant systematic errors in the high degree f modes.
We also note that the new values of $a_3$ are shifted towards higher values.

\section{Conclusion}

We analyze 15 years (1996-2011) of reprocessed MDI data and 4 years
(2010-2014) of HMI data to infer the
logarithmic radial gradient of the angular velocity of the Sun in the
upper $\sim 10$~Mm. 
By using data from two instruments and applying a different
method than CT did, we confirm their value of 
$\dd\ln\Omega/\dd\ln r \sim -1$ at low latitudes
($<30^\circ$); unlike CT, we show that $\dd\ln\Omega/\dd\ln r$ stays nearly constant and close to $-1$ up to $60^\circ$ latitude. 
With further analysis we 
conclude that the inconsistency between their results and ours for
latitudes above $30^\circ$ is due to systematic errors in the old MDI data.
This implies that work done using old MDI data
should be revisited.
By comparing the results obtained from new MDI and HMI data, we
also conclude that at least one of the data sets is likely still suffering from
some systematic errors which leads to the
discrepancy above $60^\circ$ latitude.  

The measured value $\dd\ln\Omega/\dd\ln r \sim -1$ is inconsistent with the standard
picture of angular momentum conservation where $\dd\ln\Omega/\dd\ln r$
is $-2$ \citep{Foukal77,Gilman79}.
More recently, hydrodynamical mean-field simulations of a larger part of the
convection zone by
 \cite{KiRu05} show a NSSL with a negative radial gradient of the angular
velocity from the equator to $80^\circ$ latitude.  
Their theory
\citep{KiRu93,KiRu99,Kitchatinov13}
states that the formation of the NSSL is
due to the balance of the $\Lambda$-effect \citep{Rudiger89} and the
eddy viscosity.
However, producing a NSSL with the correct radial 
gradient remains a challenge for direct numerical simulations
of the Sun \citep[e.g., ][]{Joern2013,Gustavo2013}
 and we still do not understand why the value of $\dd\ln\Omega/\dd\ln
 r$ at the surface is nearly constant and so close to $-1$. 

We note here that we measure $\dd\ln\Omega/\dd\ln r$ only in the upper
$\sim 10$~ Mm which is only about one third of the NSSL.
To extend this range one would need to use p modes, which unfortunately
have much more noise.
A preliminary analysis shows that $\dd\ln\Omega/\dd\ln r$ shows little
solar cycle variation, though there are weak hints of a torsional oscillation-like signal.
However, this requires further analysis.

\begin{acknowledgements}
We thank T.~P. Larson for discussions regarding details of the MDI and
HMI peakbagging, 
T. Corbard and M.~J. Thompson for clarifying details of their
analysis, and A. Birch 
for various discussions.
SOHO is a project of international cooperation between ESA and NASA.
The HMI data are courtesy of NASA/SDO and the HMI science team.
\end{acknowledgements}

\bibliographystyle{aa}
\bibliography{all}

\end{document}